# Photoluminescence quenching in gold - MoS$_2$ hybrid nanoflakes

*Udai Bhanu[†,⊥], Muhammad R. Islam[†,⊥], Laurene Tetard[†,⊥,*], and Saiful I. Khondaker[†,⊥,*]*
[†] NanoScience Technology Center, University of Central Florida, FL 32826, USA
[⊥] Department of Physics, University of Central Florida, FL 32826, USA
*laurene.tetard@ucf.edu; saiful@ucf.edu

**Achieving tunability of two dimensional (2D) transition metal dichalcogenides (TMDs) functions calls for the introduction of hybrid 2D materials by means of localized interactions with zero dimensional (0D) materials. A metal-semiconductor interface, as in gold (Au) - molybdenum disulfide (MoS$_2$), is of great interest from the standpoint of fundamental science as it constitutes an outstanding platform to investigate plasmonic-exciton interactions and charge transfer. The applied aspects of such systems introduce new options for electronics, photovoltaics, detectors, gas sensing, catalysis, and biosensing. Here we consider pristine MoS$_2$ and study its interaction with Au nanoislands, resulting in local variations of photoluminescence (PL) in Au-MoS$_2$ hybrid structures. By depositing monolayers of Au on MoS$_2$, we investigate the electronic structure of the resulting hybrid systems. We present strong evidence of PL quenching of MoS$_2$ as a result of charge transfer from MoS$_2$ to Au: p-doping of MoS$_2$. The results suggest new avenues for 2D nanoelectronics, active control of transport or catalytic properties.**

Recent advances in isolating single or few layers of Van der Waals (VdW) materials, such as graphene, have led to the discovery of unusual physical properties that result from 2D quantum confinement, giving rise to behavior such as quantum Hall effect [1-4], Berry phase [1,5] or ballistic carrier transport[6,7] in graphene. However, the absence of bandgap in graphene results in very low current switching and absorption (only 2.3%) of solar photons, limiting its potential for applications in electronics and optoelectronics. Recently, another VdW material, MoS$_2$, has attracted tremendous interest due to its tunable bandgap from 1.2 eV in bulk to 1.8-1.9 eV in single layers. As a result, single layer and few layers MoS$_2$ exhibit promising behavior for applications in transistor with high current switching[8,9], light emitting devices[10], photo-transistors[11-13], catalysis[14] and solar cells[15,16]. The transition to a direct bandgap in pristine MoS$_2$ monolayers[17-21], which in turns is directly related to the enhanced PL observed in monolayers, also triggered interesting work on electroluminescence[10], valley-Spin polarization[22-24], and tunable p-n diodes[25]. Other 2D layered materials and heterostructures resulting from stacking of different 2D layers are believed to hold exciting promises[26-28]. However, in most cases the production of such heterostructures is hindered by significant challenges due to the complications in obtaining other 2D materials and their control placements on top of each other. Hence, 0D-2D hybrid structures may offer a more immediate benefit for tunability of 2D TMDs functions: because layers of VdW (2D) materials are so thin, physical or chemical modifications of their surface can significantly impact their functionalities. Thus, exploring the interactions of 0D and 2D materials is crucial to the quest for highly efficient and functional hybrid nanodevices. Here, by creating hybrid 0D-2D systems, we propose to modify the properties of MoS$_2$. Hence, we investigate the effect of Au nanostructures on the electronic properties of MoS$_2$ and discuss the consequences of Au-MoS$_2$ interactions in electronics and optoelectronics.

## Results

MoS$_2$ was mechanically exfoliated[8,29,30] and transferred to a SiO$_2$ substrate (see Experimental Section and Figure 1). Figure 1 shows the optical (Figure 1a) and AFM images of pristine MoS$_2$ (Figure 1b). The optical image exhibits areas of different contrasts across the flake, which are representative of a variation from two (2L) to four layers (4L). The number of layers of the exfoliated MoS$_2$ flake was confirmed by AFM height measurements (as shown in Figure 1c) using the height profile Γ extracted from the AFM image (white Γ line) in Figure 1b. The transition from SiO$_2$ to MoS$_2$ indicates a thickness of 1.8 nm, in good agreement with previously reported thickness of two layers of MoS$_2$[31]. The number of layers in the flake was further confirmed by Raman spectroscopy (as shown in Supporting Information Figure S1) with a difference between the $E^1_{2g}$ and $A_{1g}$ peak positions of 21.5 cm$^{-1}$ for the 2L and 24.3 cm$^{-1}$ for the 4L area[32]. About 2.0 nm (thickness) of Au was then deposited by thermal evaporation (see Methods). Further AFM (Figure 1d) and Scanning Electron Microscopy (SEM) characterization (Figure 1e) of the Au-MoS$_2$ hybrid system confirmed the formation of Au nanostructures. An average height of 1.6 nm for 2.0 nm Au deposition on MoS$_2$ was measured using root mean square (RMS) data analysis (Supporting Information Figure S4) on the corresponding AFM image (Figure 1d). SEM images (Figure 1e) revealed elongated Au nanostructures, with feature size varying from 5 to 30 nm in width. AFM images indicate that the Au on MoS$_2$ exhibit some height variations within the elongated structures (Figure 1d) suggesting that they are made of several Au islands connected to each other at their base.

To investigate the effect of the Au nanostructures on the properties of MoS$_2$, we performed localized PL measurements on the pristine MoS$_2$ and Au-MoS$_2$ hybrid flake, at the same location. The PL measurements were carried out on a confocal Raman microscope system (see Experimental Section). To ensure the reproducibility of the data, we followed a careful alignment and optimization protocol. In addition, the laser power was maintained below 1 mW to avoid local heating and oxidation of the samples. The integration time was also carefully optimized to obtain a satisfactory signal-to-noise ratio while maintaining acceptable data acquisition duration and avoiding drift. Figure 2b shows the variations of the PL signal, acquired at locations labeled '2L' in Figure 2a, as a function of the selected focal plane of the excitation laser. Two emission peaks at 1.85 eV and at 1.98 eV can be seen (Figure 2b – red curve), corresponding to the A1 and B1 exciton peaks of MoS$_2$[17], whose intensities change with the focal plane of the excitation laser focus. At some focal planes, the intensity of peak B1 was higher than (f6 in Figure 2b) or equal to (f5 in Figure 2b) peak A1, while at other focal planes highest intensity for peak A1 can be obtained (f4 in Figure 2b). Such variation is usually not discussed in reported PL studies. Hence we suggest that the variation in focal plane of excitation laser should be taken into account to increase the reproducibility of 2D and hybrid systems in view of more quantitative studies. After optimization, we selected f4 for all the measurements reported in this study.

Figure 2c shows the PL spectra acquired at the same location (marked '2L' on the flake in Figure 2a) on pristine MoS$_2$ (red curve) and after deposition of 2.0 nm Au on the MoS$_2$ surface (black curve). Recent work by Splendiani et al[17] described the prominent intensity of A1 peak in PL of pristine MoS$_2$ monolayers. However A1 peak at ~1.85eV has also been observed for up to five layer (5L) MoS$_2$, although with lower amplitude than in its single layer counterpart (i.e. all parameters other than number of layers should be identical for such comparison)[17, 33, 34] The presence of A1 excitons around 1.85 eV in single layer MoS$_2$ is attributed to the direct gap transition between the maximum of the valence band and the minimum of the conduction band at the K point of the Brillouin zone in monolayers MoS$_2$. As the direct excitonic transition at the Brillouin zone K point remains constant independently of the MoS$_2$ thickness, the indirect bandgap increases as the number of MoS$_2$ layers decreases[17, 18], leading to a transition from indirect to direct gap occurs at the single layer level. The splitting between the A1 and B1 emission peaks is caused by the spin orbital and interlayer couplings and varies depending on the substrate[34]. The intensity of the B1 exciton peaks is negligible in suspended monolayer configurations and an enhancement of up to $10^4$ have been reported for A1[18] while on substrate the intensity ratio of the A1/B1 is typically less than 10 [17, 33, 35-37]. The present work was carried out on a flake made of 2L and 4L MoS$_2$ on a Si/SiO$_2$ substrate. We find that in such conditions, the intensity ratio of the A1/B1 is about 2 in 2L (Figure 2c) in close agreement with Liu et al[37]. For 4L the intensity ratio was about 1.7 and a 20% decrease of the amplitude of the peaks was measured compared to 2L. We then performed the PL measurement at the same location (marked '2L') after Au deposition. As shown in Figure 2c (black curve), both A1 and B1 excitons completely disappeared, suggesting that PL quenching has occurred.

In order to see whether the PL quenching is observed throughout the flake after Au deposition or just at a few points, we acquired PL intensity maps and peak position contour plots for A1 and B1 peaks. The maps are presented in Figure 3. Figure 3a shows PL intensity mapping of the A1 peak throughout the pristine MoS$_2$ while Figure 3c corresponds to the map of the intensity of the B1 peak. We also considered the local variation in the position of the A1 and B1 PL peaks: Figure 3e and 3g represent the contour plot of the position of peak A1 and peak B1, respectively. Interestingly, we observed local variations in PL intensity and peak position but no PL quenching at any point on the pristine MoS$_2$ flake. The intensity of peak A1 varies locally from 22 to 25 CCD Cts within the 2L region of the flake, while the intensity of peak B1 varies from 8 to 17 CCD Cts. In the 4L region, the intensity of the A1 peak varied from 16 to 18 CCD Cts. Likewise, the position of peak A1 varies locally from 1.82 eV to 1.86 eV within the flake, while the position of peak B1 varies from 1.95-2.05 eV. Recent PL studies of MoS$_2$ also show small variations in peak position for both peak A1 and peak B1[17, 33, 35, 36] and intensity ratio of A1/B1[17, 33, 35, 36] acquired using single point PL spectra. The variations could be attributed to non-homogenous interaction with the substrate, local defects in the flake or various edges configurations on the edges [33, 36-38]. Although often neglected in studies reporting single spectra studies[17, 36, 39], such 2D mappings could have a major impact on the understanding of the behavior of MoS$_2$ and defect- or hybrid-MoS$_2$. The results highlight the need for 2D mapping of the flakes to avoid the misleading observation of a sheet that would be due to a localized change in the 2D flake.

The 2D mappings were repeated after Au deposition, as presented in Figure 3b (A1 intensity), d (B1 intensity), f (A1 position), h (B1 position). No peak could be observed in the range of 1.82-1.86 eV, as can be seen in the intensity map in Figure 3b and corresponding position contour plot in Figure 3f. Similar observation holds in the range of 1.95-2.05 eV, as can be seen in the intensity map in Figure 3d and corresponding position contour plot in Figure 3h. Since Au nanostructures cover the entire MoS$_2$ surface, the observed PL quenching throughout the flake suggest strong charge transfer interaction between Au and MoS$_2$.

Recently, variations of PL intensity are reported for different substrates[34] and for liquid phase exfoliated MoS$_2$[39]. In substrate dependent studies, although no quenching was observed, the small variation attributed to the doping by the trapped charges of the substrate[34]. While for the Li-intercalated MoS$_2$ study, the absence of PL was due to a phase transition of MoS$_2$ to metallic phase[39]. In our present study, all measurements were carried out on the same substrate (Si/SiO$_2$ substrate) and the absence of PL quenching in the 2D mapping of pristine MoS$_2$ (without Au deposition) suggest that the substrate induces defects are not responsible for the observed PL quenching of $MoS_2$ after Au deposition. To determine whether the PL quenching is due to any structural change of $MoS_2$ caused by Au deposition, we performed X-ray photoemission spectroscopy (XPS) on pristine $MoS_2$ and after depositing Au. No change in the structure could be found according to a comparison of XPS peak position and peak FWHM (Supporting Information Figure S2). Thus it is clear that the PL quenching is not due to any structural change in the system.

Therefore, the variations of PL in our study suggest doping by the Au nanostructures. We infer that the PL quenching observed in the presence of Au can be attributed to a change in the electronic structure of the system. The Fermi level of a single layer of $MoS_2$ and Au was previously reported at 4.7 eV [10, 12, 40, 41] and 5.1 eV [25, 42] resulting in a 0.4 eV energy level offset (Figure 4a), which causes a band bending[12] in the Au-MoS2 hybrid (Figure 4b). Upon illumination, the electrons in the excited state of $MoS_2$ transfer to Au, leaving a hole behind, thus cause p-doping in $MoS_2$. As electrons from $MoS_2$ are transferred to Au, they do not decay back to their initial ground state, leading to PL quenching in the hybrid system.

In order to further confirm the doping of Au on $MoS_2$, we repeated the study for 1.0 nm Au deposition on another $MoS_2$ flake. The results are summarized in Figure 5. Figure 5a shows the optical image of the pristine $MoS_2$ flake considered for this sequence of measurements. The optical contrast (Figure 5a) and AFM topography (Supporting Information Figure S3) images show 2L (1.9 nm thickness) and 3L (2.4 nm thickness) areas on the flake, which was further confirmed from Raman study (Supporting Information Figure S3). The high resolution AFM image (Figure 5b) shows the morphology of the hybrid $MoS_2$ surface after 1.0 nm Au deposition with average height of 0.6 nm (extracted by RMS analysis, Supporting Information Figure S4). SEM characterization (Figure 5c) of the Au-$MoS_2$ hybrid system showed the formation of isolated triangular Au nanostructures on $MoS_2$, with size variation of 5 nm to 17 nm, corresponding to a lower coverage compared to the 2.0 nm deposition sample.

A PL spectrum of the flake was acquired (at the location indicated by the cross labeled '2L' in Figure 5a) on pristine $MoS_2$ (red curve in Figure 5d). Small variations in the intensity of both emission peaks were observed in the 2L and 3L, and at the 2L-3L transition on pristine $MoS_2$ (Figure 5e, Figure 5g and Supporting Information Figure S5). A PL spectrum of the flake at the same location was acquired after 1.0 nm Au deposition (black curve in Figure 5d). As in the 2.0 nm deposition case, we observed a significant photoluminescence quenching of the A1 peak on Au-$MoS_2$, but a small reduction of the B1. The partial quenching of the B1 peak may be due to variations in the Au nanostructures dimensions and shape. To study a local variation in the flake we acquired intensity maps of these two peaks A1 and B1. Figure 5e-h show the PL intensity maps of A1 (Figure 5e, f) and B1 (Figure 5g, h) for pristine and hybrid Au-$MoS_2$. PL quenching was observed on both 2L and 3L in presence of Au (Figure 5f and Figure 5h). The measurements confirm the p-doping of $MoS_2$ by thermally deposited Au nanostructures on 2L to 4L $MoS_2$.

Different groups have reported decoration of $MoS_2$ with Au, using chemically synthesized nanostructures[43-48]. Some of these studies presented electrical transport data, with one study showing p-doping and the other showing n-doping. Our study, on the other hand, was performed using thermally deposited Au structures, and the PL study showed p-doping. Similarly, both p- and n- doping of Au on graphene has been observed[43, 46, 49, 50]. Although the p-doping confirmed by our PL quenching result is in agreement with recent electrical transport data of $MoS_2$ [43], this quenching is rather unexpected as Au nanostructures are well known for their plasmonic characteristics and their ability to enhance materials responses[51]. Absorption around 500-600 nm has been reported for 40 nm Au nanoparticles in colloidal solutions [52] and on quartz substrates[53]. Moreover, no PL signal on the Si/SiO$_2$ substrate before Au deposition (Figure 2d (red curve), acquired at the location marked 'SiO$_2$' in Figure 2a), whereas a broad absorption centered around 2.1 eV (i.e. 590 nm) could be observed in the presence of Au nanostructures (Figure 2d (black curve)). Similar behavior was observed for 1.0 nm deposited Au (see high intensity of the substrate in Figure 5h). This suggests plasmonic activity of the Au nanostructures on Si/SiO$_2$, but did not lead to any PL enhancement on MoS$_2$. Interestingly, recent study by Sohbani et al[45], indeed shows PL enhancement on Au coated silica nanospheres of ~180 nm indicating a strong size effect of the Au nanostructures on the PL and doping of MoS$_2$. This suggests that PL and doping of MoS$_2$ by metal nanostructures is likely related to their geometry and need to be addressed in future theoretical and experimental works.

**Conclusion**

In summary, we have investigated the behavior of Au-MoS$_2$ hybrid nanoflakes resulting from Au thermal deposition. The size of the nanostructures was controlled by the thickness of Au deposited. The results show a drastic PL quenching as a result of the Au nanostructures on MoS$_2$, irrespective of the number of MoS$_2$ layers (from 2L to 4L). We suggest that the change in the electronic structure of the system consists of electrons transfer from MoS$_2$ to Au, as the work function of MoS$_2$ is lower than that of Au. Thus the electrons in the excited state of MoS$_2$ transfer to Au, leaving a hole behind, thus cause p-doping in MoS$_2$. In addition, our study demonstrates the need for more comprehensive studies of hybrid TMDs to fully understand the effect of dimension, materials, and spacing on the electronic properties of the system. However, the present results offer new options for localized tunability of the 2D TMDs properties as quenching will occur where patches of Au nanostructures are patterned. We expect these results to have numerous applications in nano-electronics and nano-optics.

**Methods**

**Few layers of MoS$_2$ were mechanically exfoliated** onto highly doped Si wafer with 250 nm thermally grown oxide layer using scotch tape. The exfoliation was performed from naturally occurring, crystalline bulk MoS$_2$, commercially available from SPI supply. Optical images to identify the MoS$_2$ flakes were acquired with an Olympus BX51 optical microscope with 100X objective.

**Raman and PL spectra** were recorded with Witec alpha300 RA confocal Raman system. The samples were illuminated with 532.0 nm laser light in ambient air environment at room temperature. The laser power was set at 0.6 mW for all data acquisition in order to avoid damage to the sample including defect formation. The measurements were collected for an integration time of 1 s. Emitted Stokes Raman signal was collected by a Zeiss 100X objective (N.A=0.9) and dispersed by a 1800 lines/mm grating for Raman measurement and a 600 lines/mm grating for PL measurements.

**Au nanostructure deposition:** Au was deposited on the mechanically exfoliated MoS$_2$ layers by thermal evaporation, at a base pressure of 1.5X10$^{-6}$ mBar at a deposition rate of 0.02 Å per second. AFM imaging and SEM images confirmed the formation of Au nanostructures. The AFM and SEM images were used to estimate the height and size of the Au nanostructures resulting from 2.0 nm and 1.0 nm deposition. AFM images were taken on a Dimension 3100 scanning probe microscope (Vecco Instruments Inc) using tapping mode. The SEM images of Au-MoS2 hybrid system was taken by Zeiss Ultra-55 SEM using Inlens detector with an accelerating voltage ∼5 kV. XPS was carried out on Physical Electronics 5400 ESCA system utilizing a monochromatized Al Kα X-ray source.

**References**

1. Zhang, Y. B., Tan, Y. W., Stormer, H. L. & Kim, P. Experimental observation of the quantum Hall effect and Berry's phase in graphene. *Nature* **438**, 201-204 (2005).
2. Novoselov, K. S., Geim, A. K., Morozov, S. V., Jiang, D., Katsnelson, M. I., Grigorieva, I. V., Dubonos, S. V. & Firsov A. A. Two-dimensional gas of massless Dirac fermions in graphene. *Nature* **438**, 197-200 (2005).
3. Jiang, Z., Zhang, Y., Tan, Y.W., Stormer, H.L. & Kim, P. Quantum Hall effect in graphene. *Solid State Commun.* **143**, 14-19 (2007).
4. Novoselov, K.S., Jiang, Z., Zhang, Y., Morozov, S. V., Stormer, H. L., Zeitler, U., Maan, J. C., Boebinger, G. S., Kim, P. & Geim, A. K. Room-Temperature Quantum Hall Effect in Graphene. *Science* **315**, 1379 (2007).
5. Carmier, P. & Ullmo, D. Berry phase in graphene: Semiclassical perspective. *Phys. Rev. B* **77**, 245413 (2008).
6. Du, X., Skachko, I., Barker, A. & Andrei, E.Y. Approaching ballistic transport in suspended graphene. *Nat. Nanotechnol.* **3**, 491-495 (2008).
7. Gunlycke, D., Lawler, H.M. & White, C.T. Room-temperature ballistic transport in narrow graphene strips. *Phys. Rev. B* **75**, 085418 (2007).
8. Radisavljevic, B., Radenovic, A., Brivio, J., Giacometti, V. & Kis, A. Single-layer $MoS_2$ transistors. *Nat. Nanotechnol.* **6**, 147-150 (2011).
9. Pu, J., Yomogida, Y., Liu, K., Li, L., Iwasa, Y. & Takenobu, T. Highly Flexible $MoS_2$ Thin-Film Transistors with Ion Gel Dielectrics. *Nano Lett.* **12**, 4013-4017 (2012).
10. Sundaram, R.S., Engel, M., Lombardo, A., Krupke, R., Ferrari, A. C., Avouris, Ph. & Steiner, M. Electroluminescence in Single Layer $MoS_2$. *Nano Lett.* **13**, 1416-1421 (2013).
11. Lee, H.S., Min, S., Chang, Y., Park, M. K., Nam, T., Kim, H., Kim, J. H., Ryu, S. & Im, S. $MoS_2$ Nanosheet Phototransistors with Thickness-Modulated Optical Energy Gap. *Nano Lett.* **12**, 3695-3700 (2012).
12. Yin, Z., Li, H., Li, H., Jiang, L., Shi, Y., Sun, Y., Lu, G., Zhang, Q., Chen, X. & Zhang, H. Single-Layer $MoS_2$ Phototransistors. *ACS Nano* **6**, 74-80 (2011).
13. Choi, W., Cho, M. Y., Konar, A., Lee, J. H., Cha, G., Hong, S. C., Kim, S., Kim, J., Jena, D., Joo, J. & Kim, S. High-Detectivity Multilayer $MoS_2$ Phototransistors with Spectral Response from Ultraviolet to Infrared. *Adv. Mat.* **24**, 5832-5836 (2012).
14. Voiry, D., Salehi, M., Silva, R., Fujita, T., Chen, M., Asefa, T., Shenoy, V. B., Eda, G. & Chhowalla, M. Conducting $MoS_2$ Nanosheets as Catalysts for Hydrogen Evolution Reaction. *Nano Lett.* **13**, 6222-6227 (2013).
15. Feng, J., Qian, X.F., Huang, C.W. & Li, J. Strain-engineered artificial atom as a broad-spectrum solar energy funnel. *Nat. Photonics* **6**, 865-871 (2012).
16. Wang, Q.H., Kalantar-Zadeh, K., Kis, A., Coleman, J.N. & Strano, M.S. Electronics and optoelectronics of two-dimensional transition metal dichalcogenides. *Nat. Nanotechnol.* **7**, 699-712 (2012).
17. Splendiani, A., Sun, L., Zhang, Y., Li, T., Kim, J., Chim, C., Galli, G. & Wang, F. Emerging Photoluminescence in Monolayer $MoS_2$. *Nano Lett.* **10**, 1271-1275 (2010).

18. Mak, K.F., Lee, C., Hone, J., Shan, J. & Heinz, T.F. Atomically Thin $MoS_2$: A New Direct-Gap Semiconductor. *Phys. Rev. Lett.* **105**, 136805 (2010).
19. Molina-Sánchez, A. & Wirtz, L. Phonons in single-layer and few-layer $MoS_2$. *Phys. Rev. B* **84**, 155413 (2011).
20. Korn, T., Heydrich, S., Hirmer, M., Schmutzler, J. & Schüller, C. Low-temperature photocarrier dynamics in monolayer $MoS_2$. *Appl. Phys. Lett.* **99**, 102109 (2011).
21. Ghatak, S., Pal, A.N. & Ghosh, A. Nature of Electronic States in Atomically Thin $MoS_2$ Field-Effect Transistors. *ACS Nano* **5**, 7707-7712 (2011).
22. Mak, K.F., He, K.L., Shan, J. & Heinz, T.F. Control of valley polarization in monolayer $MoS_2$ by optical helicity. *Nat. Nanotechnol.* **7**, 494-498 (2012).
23. Zeng, H.L., Dai, J.F., Yao, W., Xiao, D. & Cui, X.D. Valley polarization in $MoS_2$ monolayers by optical pumping. *Nat. Nanotechnol.* **7**, 490-493 (2012).
24. Xiao, D., Liu, G.-B., Feng, W., Xu, X. & Yao, W. Coupled Spin and Valley Physics in Monolayers of $MoS_2$ and Other Group-VI Dichalcogenides. *Phys. Rev. Lett.* **108**, 196802 (2012).
25. Eastman, D.E. Photoelectric Work Functions of Transition, Rare-Earth, and Noble Metals. *Phys. Rev. B* **2**, 1-2 (1970).
26. Geim, A.K. & Grigorieva, I.V. Van der Waals heterostructures. *Nature* **499**, 419-425 (2013).
27. Zhang, W., Chuu, C., Huang, J., Chen, C., Tsai, M., Chang, Y., Liang, C., Chen, Y., Chueh, Y., He, J., Chou, M. & Li, L. Ultrahigh-Gain Photodetectors Based on Atomically Thin Graphene-$MoS_2$ Heterostructures. *Sci. Rep.* **4,** 3826 (2014).
28. Eda, G., Fujita, T., Yamaguchi, H., Voiry, D., Chen, M. & Chhowalla M. Coherent Atomic and Electronic Heterostructures of Single-Layer $MoS_2$. *ACS Nano* **6**, 7311-7317 (2012).
29. Novoselov, K.S., Jiang, D., Schedin, F., Booth, T. J., Khotkevich, V. V., Morozov, S. V. & Geim A. K. Two-dimensional atomic crystals. *Proc. Natl. Acad. Sci. USA* **102**, 10451-10453 (2005).
30. Brivio, J., Alexander, D.T.L. & Kis, A. Ripples and Layers in Ultrathin $MoS_2$ Membranes. *Nano Lett.* **11**, 5148-5153 (2011).
31. Huang, X., Zeng, Z.Y. & Zhang, H. Metal dichalcogenide nanosheets: preparation, properties and applications. *Chem. Soc. Rev.* **42**, 1934-1946 (2013).
32. Lee, C., Yan, H., Brus, L. E., Heinz, T. F., Hone J. & Ryu, S. Anomalous Lattice Vibrations of Single- and Few-Layer $MoS_2$. *ACS Nano* **4**, 2695-2700 (2010).
33. Mouri, S., Miyauchi, Y. & Matsuda, K. Tunable Photoluminescence of Monolayer $MoS_2$ via Chemical Doping. *Nano Lett.* **13**, 5944-5948 (2013).
34. Buscema, M., Steele, G., Zant, H.J. & Castellanos-Gomez, A. The effect of the substrate on the Raman and photoluminescence emission of single-layer $MoS_2$. *Nano Res.* **7**, 1-11 (2014).
35. Tongay, S., Suh, J. Ataca, C. Fan, W., Luce W., Kang, J. S., Liu, J., Ko, C., Raghunathanan, R., Zhou, J., Ogletree, F., Li, J., Grossman J. C., & Wu, J. Defects activated photoluminescence in two-dimensional semiconductors: interplay between bound, charged, and free excitons. *Sci. Rep.* **3,** 2657 (2013).
36. Scheuschner, N., Ochedowski, O., Kaulitz, A., Gillen, R., Schleberger, M. & Maultzsch J. Photoluminescence of freestanding single- and few-layer $MoS_2$. *Phys. Rev. B* **89**, 125406 (2014).

37. Liu, Y.L., Nan, H., Wu, X., Pan, W., Wang, W., Bai. J., Zhao, W., Sun, L., Wang, X. & Ni, Z. Layer-by-Layer Thinning of $MoS_2$ by Plasma. *ACS Nano* **7**, 4202-4209 (2013).
38. Su, L., Zhang, Y., Yu, Y. & Cao, L. Dependence of coupling of quasi 2-D $MoS_2$ with substrates on substrate types, probed by temperature dependent Raman scattering. *Nanoscale* **6**, 4920-4927 (2014).
39. Eda, G., Yamaguchi, H., Voiry, D., Fujita, T., Chen, M. & Chhowalla M. Photoluminescence from Chemically Exfoliated $MoS_2$. *Nano Lett.* **11**, 5111-5116 (2011).
40. Liu, K.-K., Zhang, W., Lee, Y., Lin, Y., Chang, M., Su, C., Chang, C., Li, H., Shi, Y., Zhang, H., Lai, C. & Li, L. Growth of Large-Area and Highly Crystalline $MoS_2$ Thin Layers on Insulating Substrates. *Nano Lett.* **12**, 1538-1544 (2012).
41. Choi, M.S., Lee, G., Yu,Y., Lee, D., Lee, S. W., Kim, P., Hone, J. & Yoo, W. J. Controlled charge trapping by molybdenum disulphide and graphene in ultrathin heterostructured memory devices. *Nat. Commun.* **4**, 1624 (2013).
42. Michaelson, H.B. The work function of the elements and its periodicity. *J. Appl. Phys.* **48**, 4729-4733 (1977).
43. Shi, Y., Huang, J., Jin, L., Hsu, Y., Yu, S. F., Li, L. & Yang, H. Y. Selective Decoration of Au Nanoparticles on Monolayer $MoS_2$ Single Crystals. *Sci. Rep.* **3**, 1839 (2013).
44. Sreeprasad, T.S., Nguyen, P., Kim, N. & Berry, V. Controlled, Defect-Guided, Metal-Nanoparticle Incorporation onto $MoS_2$ via Chemical and Microwave Routes: Electrical, Thermal, and Structural Properties. *Nano Lett.* **13**, 4434-4441 (2013).
45. Sobhani, A., Lauchner, A., Najmaei, S., Ayala-Orozco, C., Wen, F., Lou, J. & Halas N. J. Enhancing the photocurrent and photoluminescence of single crystal monolayer $MoS_2$ with resonant plasmonic nanoshells. *Appl. Phys. Lett.* **104**, 031112 (2014).
46. Shi, Y., Kim, K. K., Reina, A., Hofmann, M., Li, L. & Kong, J. Work Function Engineering of Graphene Electrode via Chemical Doping. *ACS Nano* **4**, 2689-2694 (2010).
47. Scharf, T.W., Goeke, R.S., Kotula, P.G. & Prasad, S.V. Synthesis of Au–$MoS_2$ Nanocomposites: Thermal and Friction-Induced Changes to the Structure. *ACS Appl. Mater. Inter.* **5**, 11762-11767 (2013).
48. Rao, B.G., Matte, H. S.S. R., Rao, C.N.R. Decoration of Few-Layer Graphene-Like MoS2 and MoSe2 by Noble Metal Nanoparticles. *J. Clust. Sci.* **23**, 929-937 (2012).
49. Ren, Y., Chen, S., Cai, W., Zhu, Y., Zhu, C. & Ruoff, R. S. Controlling the electrical transport properties of graphene by in situ metal deposition. *Appl. Phys. Lett.* **97**, 053107 (2010).
50. Huh, S., Park, J., Kim, K.S., Hong, B.H. & Kim, S.B. Selective n-Type Doping of Graphene by Photo-patterned Gold Nanoparticles. *ACS Nano* **5**, 3639-3644 (2011).
51. Jasuja, K. & Berry, V. Implantation and Growth of Dendritic Gold Nanostructures on Graphene Derivatives: Electrical Property Tailoring and Raman Enhancement. *ACS Nano* **3**, 2358-2366 (2009).
52. Aslan, K., Holley, P., Davies, L., Lakowicz, J.R. & Geddes, C.D. Angular-Ratiometric Plasmon-Resonance Based Light Scattering for Bioaffinity Sensing. *J. Am. Chem. Soc.* **127**, 12115-12121 (2005).
53. Lereu, A. L., Farahi, R. H., Tetard L., Enoch, S., Thundat, T. & Passian A. Plasmon assisted thermal modulation in nanoparticles. *Opt. Express* **21**, 12145-12158 (2013).

**Additional Information**

*Author contribution statement*
U.B. and M.R.I. prepared the samples and performed the XPS, optical and SEM imaging. U.B. performed AFM imaging. U.B. and L.T. acquired the PL and Raman spectroscopy data. L.T. and S.K. directed the experiment. U.B., L.T. and S.K. contributed to the preparation of the manuscript. All authors reviewed the manuscript.

*Competing financial interests*
The authors declare no competing financial interests.
9

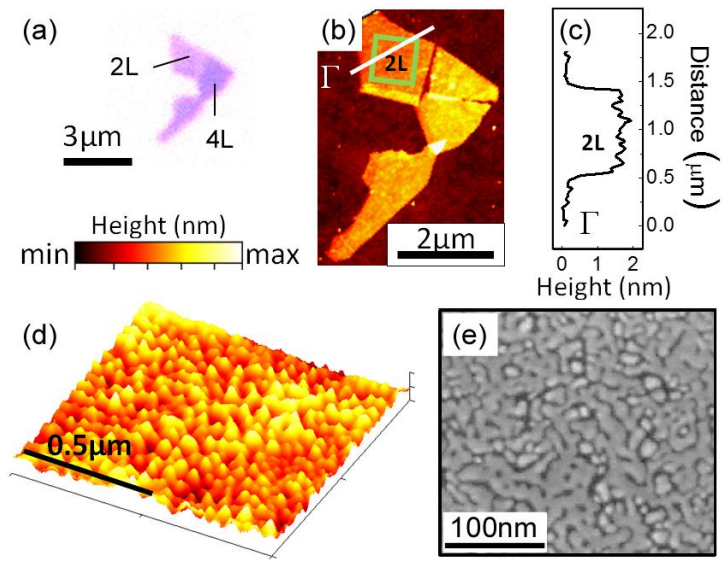

**Figure 1: Exfoliated MoS2 flake and 2.0 nm Au deposition** (a) Optical image of exfoliated MoS$_2$ flake. Different colors correspond to the different number of MoS$_2$ layers: 2 layers (2L) and 4 layers (4L). (b) AFM image of the MoS$_2$ flake as exfoliated (color scale: min=10.4nm, max=18.4nm). Γ line (black) corresponds to the profile cross-section extracted (c) to confirm the height of the two layer. (d) High resolution AFM image Au nanoislands on MoS$_2$ corresponding to boxed area (green) in (b) (color scale: min=0.0nm, max=2.9nm) (e) SEM image of Au nanostructures on MoS$_2$.

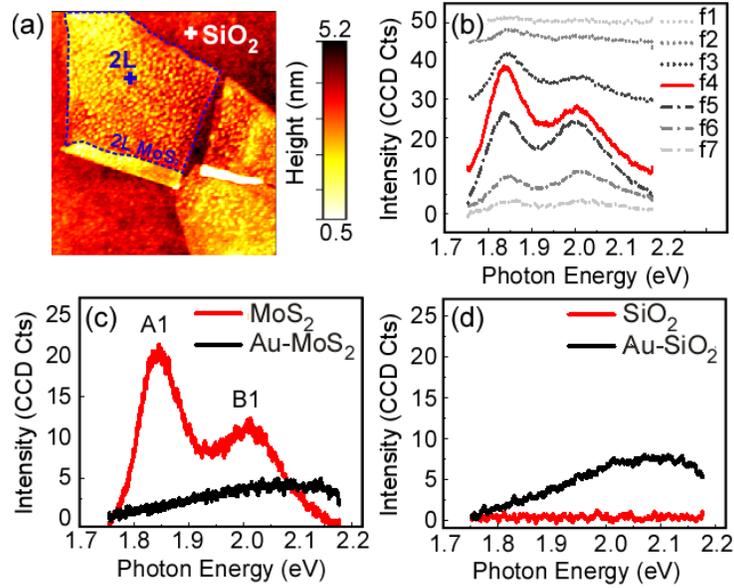

**Figure 2: Effect of 2.0 nm Au deposition on MoS2 photoluminescence (PL).** (a) AFM image of $MoS_2$ flake as exfoliated. Cross marks indicate the location of the PL data presented in (b). (b) PL signal as a function of focal plane from above the surface (f1) to several tens of nm below the surface (f7). For clarity the curves were shifted of 4 units for each step. (c) PL spectra of 2L $MoS_2$ flake as exfoliated (red) and after 2.0 nm Au deposition (black). (d) Plasmonic response of the bare $SiO_2$ substrate (red) and $SiO_2$ substrate covered with the Au nanoislands (black).

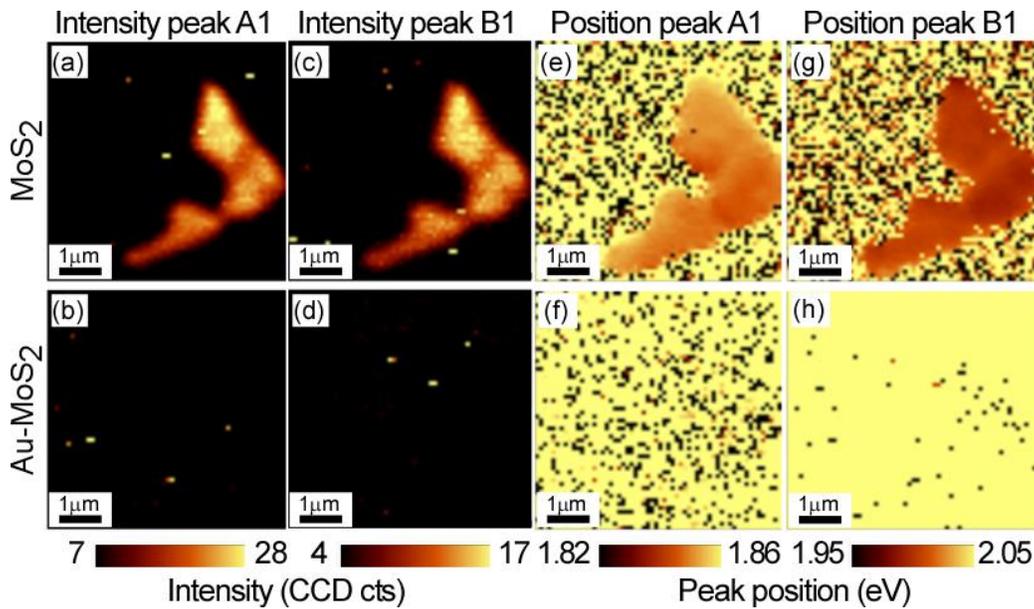

**Figure 3: Photoluminescence (PL) mapping of Au-MoS$_2$ hybrid nanoflake** Intensity (a) and position (e) of peak A1 across the exfoliated MoS$_2$ flake, Intensity (b) and position (f) of peak A1 across the exfoliated Au-MoS$_2$ flake (after 2.0 nm Au deposition). Intensity (c) and position (g) of peak B1 across the exfoliated MoS$_2$ flake, Intensity (d) and position (h) of peak B1 across the exfoliated Au-MoS$_2$ flake (after 2.0 nm Au deposition).

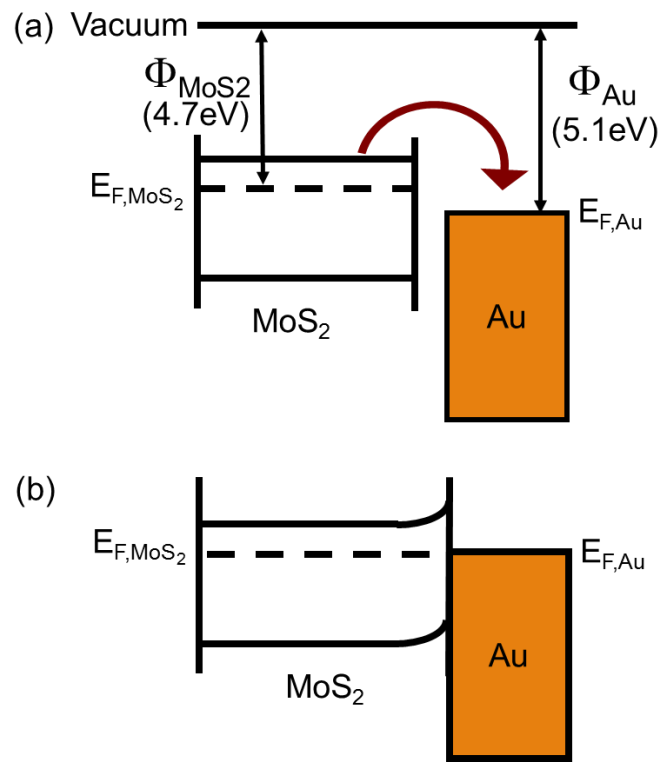

**Figure 4: Electron transfer mechanism in Au-MoS$_2$ hybrid nanoflake.** (a) The energy band diagram for MoS$_2$ and Au shows the relative positions of Fermi level with respect to vacuum level before establishing a contact. The direction of the arrow represents the transfer of electrons from MoS$_2$ to Au after the contact is established, (b) The energy band diagram of Au-MoS$_2$ showing band bending after establishing the contact between Au and MoS$_2$. Electron transfer from MoS$_2$ to Au causes p-doping and PL quenching.

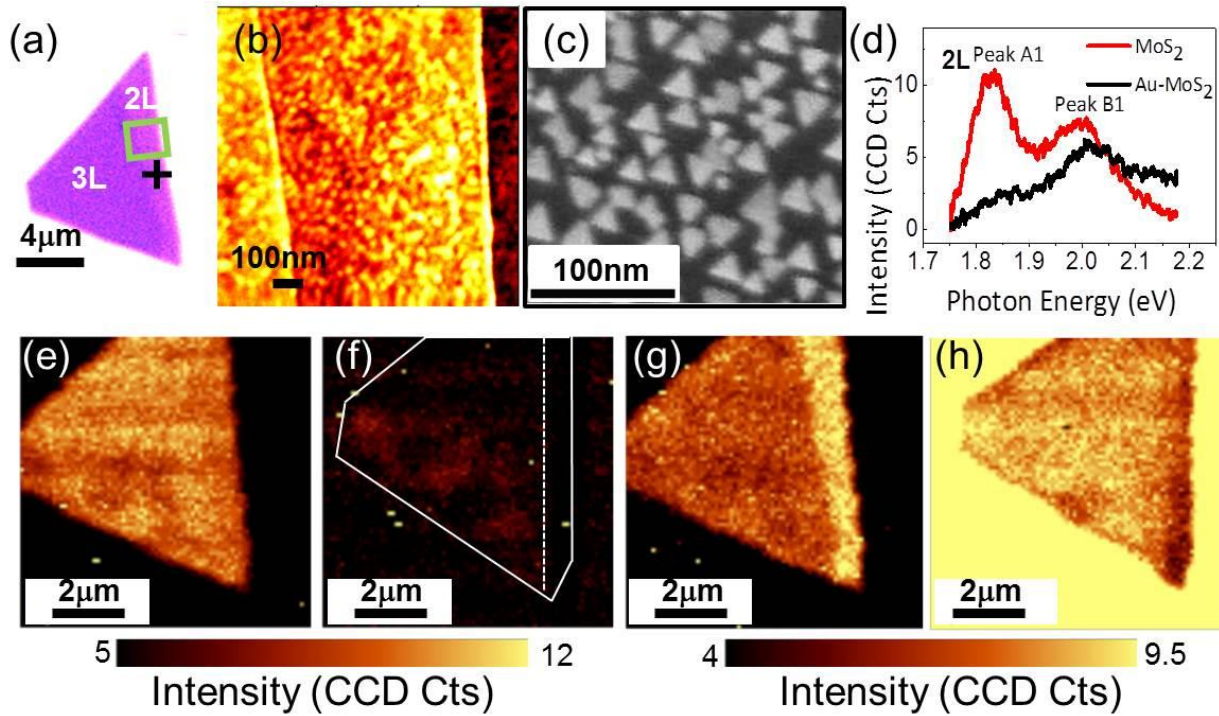

**Figure 5: Effect of 1.0 nm Au evaporation on the photoluminescence properties of exfoliated MoS$_2$.** (a) Optical image of exfoliated MoS$_2$ flake. Different color contrasts show the different number of layers, corresponding to 2 layers (2L) and 3 layers (3L) as indicated on the image. (b) AFM image of MoS$_2$ flake after 1.0 nm Au deposition. (c) SEM image of MoS$_2$ flake after 1.0 nm Au deposition. (d) PL curve of the pristine MoS$_2$ flake (red) and Au-MoS$_2$ hybrid after 1.0 nm Au deposition (black) acquired at crossed labeled 2L in (a), showing PL quenching of peak A1 and B1. PL intensity maps of peak A1 (e, f) and B1 (g,h) before (e,g) and after (f,h) 1.0 nm Au deposition. Contours were used in (f) to facilitate the comparison.

# Supplementary information

# Photoluminescence quenching in gold - MoS$_2$ hybrid nanoflakes

*Udai Bhanu*[†,⊥], *Muhammad R. Islam*[†,⊥], *Laurene Tetard*[†,⊥,*], *and Saiful I. Khondaker*[†,⊥,*]

[†] NanoScience Technology Center, University of Central Florida, FL 32826, USA
[⊥] Department of Physics, University of Central Florida, FL 32826, USA
*laurene.tetard@ucf.edu; saiful@ucf.edu

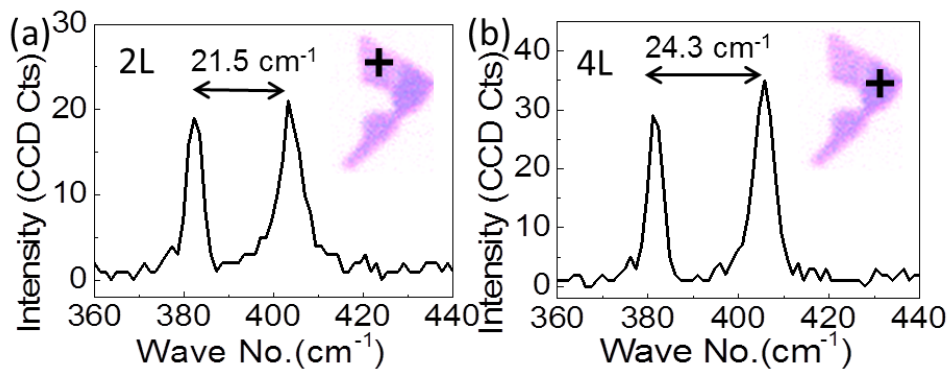

**Figure S1**. Raman Spectrum of the two layer (2L) and 4 layer (4L) of pristine $MoS_2$ flake. The difference between $E_{2g}$ and $A_{1g}$ peak is 21.5 and 24.3 cm$^{-1}$, respectively. The difference in wave number measured corresponds to presence of 2L and 4L in the $MoS_2$ flake. Inset shows the optical micrograph of the $MoS_2$ flake with cross marks indicating the location for Raman spectrum acquisition.

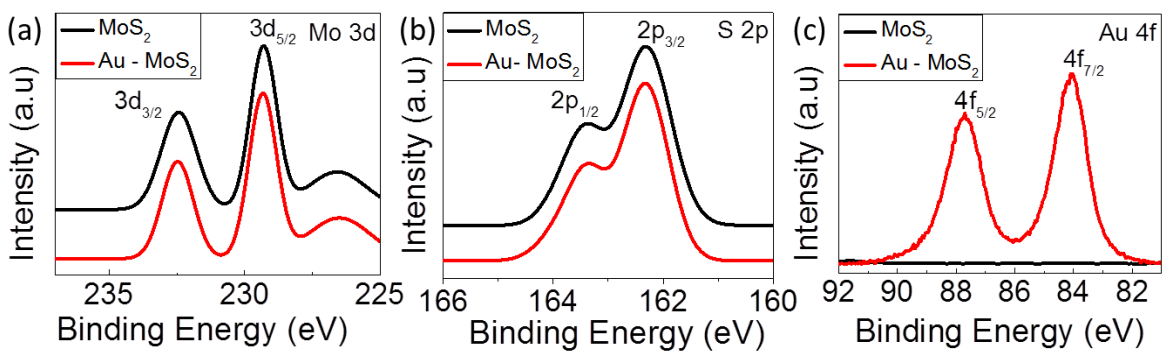

**Figure S2:** X-Ray Photoemission Spectroscopy (XPS) analysis of pristine $MoS_2$ (black curve) and Au-$MoS_2$ hybrid (red curve) for (a) Mo with Mo 3d binding energies peaks at 229.3 and 232.5 eV (b) S with S 2p binding energies peaks at 162.3 and 163.4 eV and (c) Au with Au 4f binding energies peaks at 84 and 87.8 eV. XPS data reveals that the position of Mo 3d and S 2p peaks does not change after Au deposition indicating that Au nanostructures do not change the crystal structure of $MoS_2$ flake.

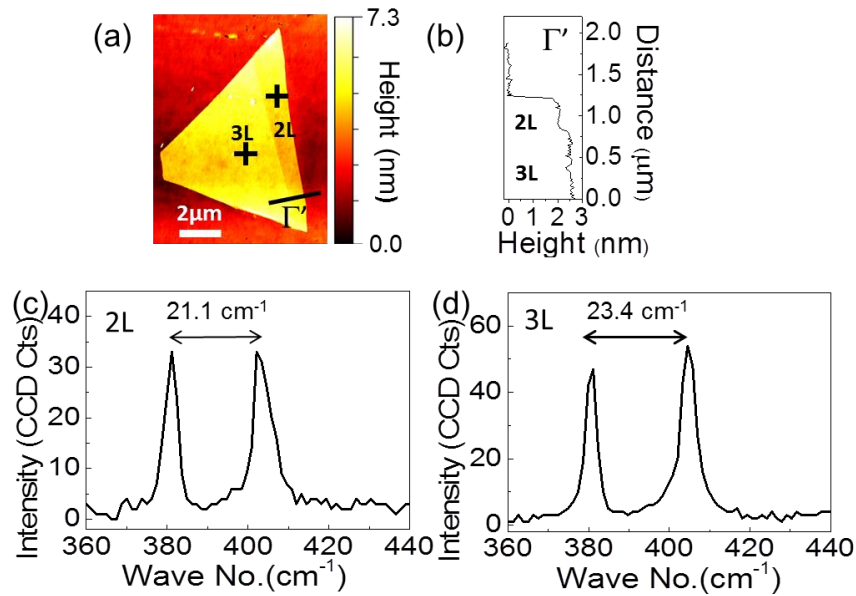

**Figure S3:** (a) AFM image of the pristine MoS$_2$ flake used for 1 nm Au deposition study. Γ' line (black) indicates the position at which the height profile cross-section (b) was extracted to measure the height of the flake MoS$_2$ layers. (c,d) Raman spectra at location labeled '2L' and '3L' in (a). The difference between E$_{2g}$ and A$_{1g}$ peak were measured: 21.1 (c) and 23.4 cm$^{-1}$ (d), which is in agreement with the height for 2L and 3L region in the MoS$_2$ flake.

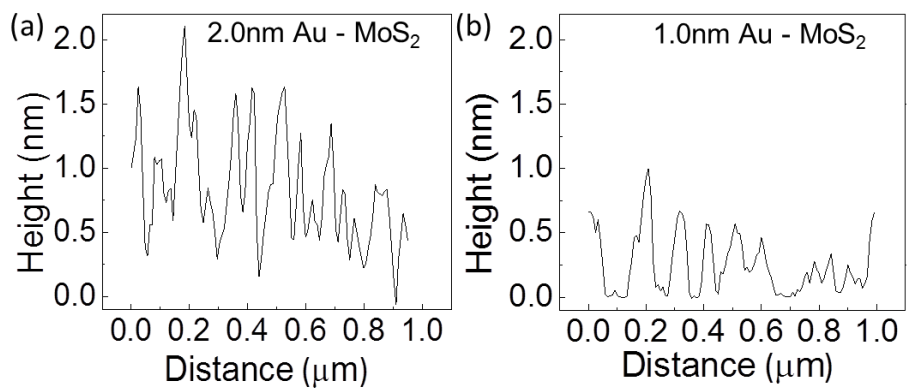

**Figure S4.** Height profiles used for Root mean square (RMS) analysis on MoS$_2$ flakes after 2.0 nm (a) and 1.0 nm (b) Au deposition on MoS$_2$. Average heights of Au nanostructures were found to be 1.6 nm for 2.0 nm deposition (a) and 0.6 nm for 1.0 nm deposition (b).

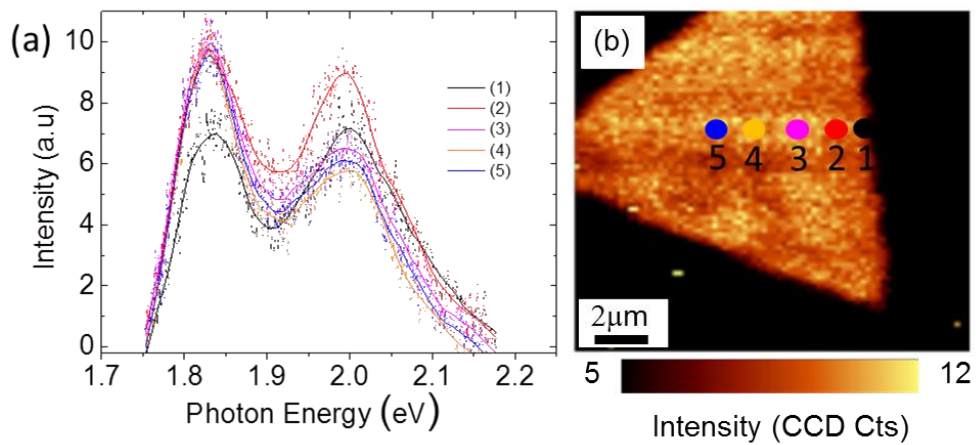

**Figure S5.** (a) Local variation in the PL intensity of pristine MoS2 flake used for 1 nm Au deposition study. Spectra were taken at five different points along a line as shown in the intensity map of A1 peak (b). (1) in black shows the signal acquired at 2L close to the edge of the nanoflake, while (2) in red shows the PL signature of 2L closer to the 3L interface. (3)-(5) are representative of the 3L $MoS_2$ PL signature.